\documentclass[amsmath,amssymb,amsfonts,twocolumn]{revtex4}
\usepackage{graphicx}
\usepackage{bbm}
\usepackage{braket}
\def \beq {\begin{equation}}
\def \eeq {\end{equation}}
\def \tr {\rm Tr}
\newcommand {\qs} {{\rm Q}_{\rm S}}
\newcommand {\qt} {{\rm Q}_{\rm T}}
\newcommand {\ks} {k_{\rm S}}
\newcommand {\kt} {k_{\rm T}}

\begin{document}
\title{Quantum trajectory tests of radical-pair quantum dynamics in CIDNP measurements of photosynthetic reaction centers}
\author{K. Tsampourakis and I. K. Kominis}
\email{ikominis@physics.uoc.gr}

\affiliation{Department of Physics, University of Crete, Heraklion
71103, Greece}

\begin{abstract}
Chemically induced dynamic nuclear polarization is a ubiquitous phenomenon in photosynthetic reaction centers. The relevant nuclear spin observables are a direct manifestation of
the radical-pair mechanism. We here use quantum trajectories to describe the time evolution of radical-pairs, and compare their prediction of nuclear spin observables to the one derived from the radical-pair master equation. While our approach provides a consistent description, we unravel a major inconsistency within the conventional theory, thus challenging the theoretical interpretation of numerous CIDNP experiments sensitive to radical-pair reaction kinetics.\end{abstract}

\maketitle
\section{Introduction}
Quantum coherence in the light-harvesting process of photosynthesis \cite{aspuru,ishizaki,scholes,mizel,fleming,scholes_review,brumer,engel,plenio,rozzi,renger,coker,collini,OC,thorwart} is a central theme in the growing field of quantum biology \cite{plenio_review}. Equally important for understanding photosynthesis and potentially designing biomimetic devices harvesting solar energy is the charge separation following the trapping of the exciton's energy in the photosynthetic reaction center. Charge separation proceeds via a cascade of electron transfers between radical-ion pairs, as for example bacteriochlorophyll and bacteriopheophytin. In parallel with charge, spin also plays a ubiquitous role \cite{matysik_review}, with the effect of chemically induced dynamic nuclear polarization (CIDNP) manifested in a wide range of photosynthetic reaction centers \cite{closs,kaptein,beyerle,boxer,zys,mcdermott,prakash,diller,daviso,matysik_pnas,matysik_PR,jeschke_TSM,JM,jeschke_lowB}. 

Nuclear spin effects in CIDNP are regulated by the radical-pair mechanism, which describes a class of spin-dependent chemical reactions studied by spin chemistry \cite{haberkorn,steiner1,steiner2,ritz2000,woodward,rodgers_review,hore_PNAS}. We have recently addressed \cite{pre2009,pre2011,pre2012,cidnp,lamb,pre2014} the fundamental quantum dynamics of the radical-pair mechanism, using concepts from quantum information science to describe the intertwined effects of coherent spin motion and spin-dependent charge recombination of radical-pairs, arriving at what we understand is a fundamental master equation describing the time evolution of $\rho$, the radical-pair's spin density matrix. The traditional (also called Haberkorn) theory \cite{haberkorn} used  until now is a limiting case of our theory valid in the regime of strong spin relaxation.

Interestingly, in CIDNP measurements the lifetimes of singlet and triplet radical-pairs (RPs) are small enough to allow observation of quantum effects without them being masked by spin relaxation. Hence CIDNP appears to be an ideal setting to test our understanding of the quantum dynamics of the radical-pair mechanism. To this end, we demonstrated \cite{cidnp} a notable difference between our theory and Haberkorn's approach in predicting a CIDNP effect at earth's field. However, measurements at earth's field are still challenging. Moreover, in \cite{cidnp} we assumed equal singlet and triplet recombination rates, $\ks=\kt$, whereas in real reaction centers we encounter asymmetric recombination rates, $\kt\gg\ks$ \cite{matysik_PR}.

Instead of attempting a comparison of the two different approaches based on an absolute quantitative prediction, we here use CINDP as a testbed to study {\it the internal consistency} of the two theories, Haberkorn's and ours, and we do so using the {\it realistic} (i.e. asymmetric) recombination rates and high magnetic fields pertinent to CIDNP experiments. It is a basic fact of the theory of open quantum systems that the time evolution of an ensemble of systems described by a master equation must exactly reproduce the average calculated from single-system realizations, called quantum trajectories. We here use a central CIDNP observable, the nuclear spin polarization of the radical-pair's reaction products, and show that Haberkorn's theory produces vastly different predictions when using Haberkorn's master equation compared to averaging Haberkorn's quantum trajectories, while our theory produces largely consistent predictions either way. 
\section{Quantum measurement approach to radical-pair quantum dynamics}
\subsection{Master equation approach}
\begin{figure}[t!]
\begin{center}
\includegraphics[width=8 cm]{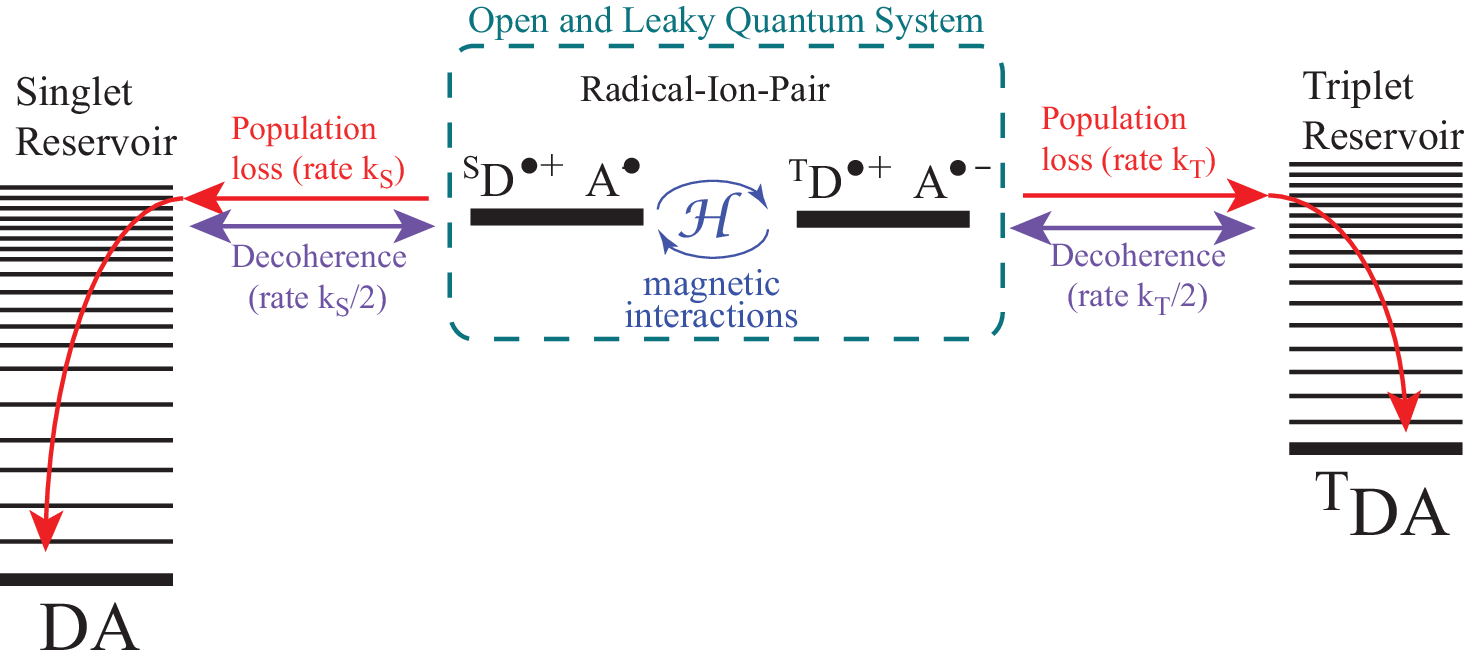}
\caption{A charge transfer from a photoexcited donor-acceptor dyad DA produces a radical-ion pair in the singlet state $^{\rm S}{\rm D}^{\bullet +}{\rm A}^{\bullet -}$. Intramolecule magnetic interactions (Hamiltonian ${\cal H}$) induce a coherent interconversion between singlet and triplet radical-pairs, $^{\rm S}{\rm D}^{\bullet +}{\rm A}^{\bullet -}~\leftrightharpoons~^{\rm T}{\rm D}^{\bullet +}{\rm A}^{\bullet -}$, terminated by the recombination event, that results into a singlet, DA, or a triplet neutral product, $^{\rm T}$DA. Radical-pairs are leaky systems because real transitions to the singlet and triplet vibrational reservoir states lead to population loss at the recombination rate $\ks$ and $\kt$, respectively. They are also open quantum systems, because virtual transitions to the same reservoir states and back to the radical-pair lead to singlet-triplet dephasing at the rate $(\ks+\kt)/2$.}
\label{rp}
\end{center}
\end{figure}

In recent years we have addressed the fundamental quantum dynamics of radical-pair reactions, schematically depicted in Fig.\ref{rp}. This biochemical system is both a leaky {\it and} an open quantum system, losing population due to the spin-dependent recombination reactions, while simultaneously suffering decoherence. The latter aspect of the dynamics was addressed in \cite{pre2009}, and from a slightly different perspective in \cite{lamb}, using quantum measurement theory. The former was phenomenologically considered in  \cite{pre2011}, while a first-principles approach was developed in \cite{pre2014}, where we formally defined a singlet-triplet coherence measure, called $p_{\rm coh}$. We also utilized the quantum-communications concept of quantum retrodiction to derive the spin-dependent reaction terms of the master equation. We will briefly recapitulate the results of \cite{pre2014} for completeness of this work.

The full master equation we arrived at in \cite{pre2014}, the consistency of which we will explore in this work, reads
\begin{align}
{{d\rho}\over {dt}}=&-i[{\cal H},\rho]\label{t1}\\  
&-{{k_{\rm S}+k_{\rm T}}\over 2}\big(\rho {\rm {\rm Q_S}}+{\rm {\rm Q_S}}\rho-2{\rm {\rm Q_S}}\rho {\rm {\rm Q_S}}\big)\label{t2}\\
&-(1-p_{\rm coh})\big(k_{\rm S}{\rm {\rm Q_S}}\rho {\rm {\rm Q_S}}+k_{\rm T} {\rm Q_T} \rho {\rm Q_T}\big)\label{t3}\\
&-p_{\rm coh}{{dr_{\rm S}+dr_{\rm T}}\over {dt}}{1\over {\tr\{\rho\}}}
\Big(\qs\rho\qs+\qt\rho\qt+\nonumber\\&\hspace{1in}{1\over p_{\rm coh}}\qs\rho\qt+{1\over p_{\rm coh}}\qt\rho\qs\Big)\label{t4}
\end{align}
The density matrix $\rho$ describes the spin state of the radical-ion-pair, consisting of the two unpaired electrons and the nuclear spins residing in the two radicals. The density matrix and all other operators are represented by $d\times d$ matrices, where $d$ is the dimension of the total spin space of the radical-pair. In this work we will consider a radical-pair having just one nuclear spin, hence $d=8$, and the 8 basis kets are $\ket{\rm S}\otimes\ket{\Uparrow}$, $\ket{\rm S}\otimes\ket{\Downarrow}$, $\ket{\rm T_+}\otimes\ket{\Uparrow}$, $\ket{\rm T_+}\otimes\ket{\Downarrow}$, $\ket{\rm T_0}\otimes\ket{\Uparrow}$, $\ket{\rm T_0}\otimes\ket{\Downarrow}$, $\ket{\rm T_-}\otimes\ket{\Uparrow}$, $\ket{\rm T_-}\otimes\ket{\Downarrow}$. The two-electron state is on the left of the tensor product and the nuclear spin state on the right. The two-electron spin subspace is spanned by the singlet $\ket{\rm S}=(\ket{\uparrow\downarrow}-\ket{\downarrow\uparrow})/\sqrt{2}$ and the triplets $\ket{\rm T_+}=\ket{\uparrow\uparrow}$, $\ket{\rm T_0}=(\ket{\uparrow\downarrow}+\ket{\downarrow\uparrow})/\sqrt{2}$, $\ket{\rm T_-}=\ket{\downarrow\downarrow}$.

The operators $\qs$ and $\qt$ project the radical-pair spin state onto the electron singlet and triplet subspace, respectively. They are orthogonal, complete and idempotent, i.e. $\qs\qt=\qt\qs=0$, $\qs+\qt=\mathbbmtt{1}$, where $\mathbbmtt{1}$ is the $d\times d$ unit matrix, and $\qs^2=\qs$, $\qt^2=\qt$. Writing $\mathbf{s}_{D}$ and $\mathbf{s}_{A}$ for the spin operator of the donor's and acceptor's unpaired electron, respectively, it is $\qs={1\over 4}\mathbbmtt{1}-\mathbf{s}_{D}\cdot\mathbf{s}_{A}$. 

The rates $\ks$ and $\kt$ are the singlet and triplet recombination rates. If at $t=0$ we prepare an RP ensemble in the singlet (triplet) electron spin state, and assume that there is no singlet-triplet (S-T) mixing, then the RP population would decay exponentially at a rate $\ks$ ($\kt$). These two rates are properties of the particular RP under consideration. In principle they can be calculated from electron transfer theory, but in practice they are determined from experiment. There are RPs for which $\ks=\kt$, and there are RPs for which $\ks\neq\kt$.

The term \eqref{t1} in the previous master equation is the ordinary unitary evolution driven by the intramolecule magnetic interactions contained in the Hamiltonian ${\cal H}$ (Zeeman, hyperfine etc). The particular Hamiltonian used in this work is relevant to CINDP measurements and will be outlined in Section IV. Since singlet and triplet states are not eigenstates of ${\cal H}$, the term \eqref{t1} generates S-T coherence. 

This is dissipated by the Lindblad term \eqref{t2}, which we derived in \cite{pre2009,lamb}, and which describes a continuous quantum measurement of $\qs$ performed by the vibrational reservoirs of Fig.\ref{rp}, the results of which are unobserved. This measurement effects projections to the electron singlet or triplet subspace suffered by individual RPs at random times. The interruptions of the coherent S-T mixing at the single-molecule level lead to S-T decoherence at the ensemble level. 

The final two terms \eqref{t3} and \eqref{t4} are the so-called reaction terms, reducing the RP population, given by $\tr\{\rho\}$, in a spin-dependent way. The normalization of the initial density matrix is $\tr\{\rho_0\}=1$. If we choose a small enough $dt$ so that $\ks dt,\kt dt\ll 1$, the fraction of the RP population that will recombine into singlet and triplet neutral products within the interval $dt$ around time $t$ is $dr_{\rm S}=\ks dt\tr\{\rho\qs\}$ and $dr_{\rm T}=\kt dt\tr\{\rho\qt\}$, respectively. Based on how coherent is the RP ensemble at time $t$, quantified by $p_{\rm coh}$, which is a function of $\rho$ straightforward to calculate \cite{pre2014}, we use the theory of quantum retrodiction to probabilistically estimate the pre-recombination state of the observed reaction products and thus arrive at \eqref{t3} and \eqref{t4}. As expected, it is $d\tr\{\rho\}=-dr_{\rm S}-dr_{\rm T}$. In the single-molecule picture, $dr_{\rm S}$ ($dr_{\rm T}$) is the probability that a {\it single} radical-pair will recombine in the singlet (triplet) channel during the time interval $dt$.

The values of $p_{\rm coh}$ range from 0, describing a maximally incoherent mixture of singlet and triplet RPs, to $0<p_{\rm coh}<1$ describing partially coherent, to $p_{\rm coh}=1$ describing maximally coherent RPs. When $p_{\rm coh}=0$, the whole term \eqref{t4} vanishes.

Given (i) the two rates $\ks$ and $\kt$, (ii) the specific Hamiltonian ${\cal H}$, and (iii) the initial state $\rho_0$, Eq. \eqref{t1}-\eqref{t4} can be used to propagate $\rho$ in time and thus calculate the time-dependence of any physical observable of interest. 
\subsection{Quantum trajectory approach}
The new physics of RP quantum dynamics that we introduced in \cite{pre2009,pre2011} is that during its lifetime, a radical-pair undergoes random projections to the singlet or triplet electron spin subspace. These projections take place at different and random times for each radical-pair. They run simultaneously with and independently of the second kind of random event, the RP charge recombination, which terminates the reaction. To generate quantum trajectories for an RP being in the state $|\psi\rangle$ at time $t$, we thus have to consider in total 5 possible events that can take place in the following time interval $dt$:
\begin{table}[h!]
    \centering
    \caption{Quantum trajectory time evolution according to Kominis' approach.}
    \begin{tabular}{|c|c|c|}
        \hline
        Name & Event & Probability of event\\\hline\hline
        K1&projection to the singlet  &$dp_{\rm S}=dt{{\ks+\kt}\over 2}\langle\psi|\qs|\psi\rangle$\\
            & state ${{\qs|\psi\rangle}\over \sqrt{\langle\psi|\qs|\psi\rangle}}$&\\\hline
        K2&projection to the triplet  &$dp_{\rm T}=dt{{\ks+\kt}\over 2}\langle\psi|\qt|\psi\rangle$\\
            & state ${{\qt|\psi\rangle}\over \sqrt{\langle\psi|\qt|\psi\rangle}}$&\\\hline
        K3 & singlet recombination & $dr_{\rm S}=\ks dt\langle\psi|\qs|\psi\rangle$\\\hline
        K4 & triplet recombination & $dr_{\rm T}=\kt dt\langle\psi|\qt|\psi\rangle$\\\hline
        K5 & hamiltonian evolution & $1-dp_{\rm S}-dp_{\rm T}-dr_{\rm S}-dr_{\rm T}$\\\hline
    \end{tabular}
\end{table}

It is well known from the theoretical treatment of open quantum systems \cite{petruccione} that the master equation describing the time evolution of the system's density matrix should exactly reproduce the average of many single-system quantum trajectories.
Thus, the average of many trajectories formed by K1-K5 should exactly reproduce the results of the master equation \eqref{t1}-\eqref{t4}. We will check whether this is the case in the context of CIDNP observables in Section IV.
\section{Haberkorn approach to radical-pair quantum dynamics}
\subsection{Master equation approach}
The traditional, or Haberkorn master equation reads 
\beq
{{d\rho}\over {dt}}=-i[{\cal H},\rho]-{\ks\over 2}(\qs\rho+\rho\qs)-{\kt\over 2}(\qt\rho+\rho\qt)\label{hab}
\eeq
Interestingly, Haberkorn's master equation follows from our master equation \eqref{t1}-\eqref{t4} by forcing $p_{\rm coh}$ to be zero at all times. In any case, as we have done for our approach in the previous section, Haberkorn's approach must be able to provide the equivalent picture of single-molecule quantum trajectories. However,  the concept of quantum trajectories has not been utilized in spin chemistry so far. Furthermore, although a recent experiment \cite{molin} provided evidence for the physical reality of the S-T decoherence process we introduced, a general consensus on what exactly is the quantum state evolution of surviving RPs is still missing. From our perspective, we do not see how Haberkorn's approach, being phenomenological, can incorporate quantum trajectories without going into the derivations of \cite{pre2009,lamb}, which lead to events K1 and K2, but we leave it as an open question to be addressed by the proponents of the conventional theory. Nevertheless, we will here outline some rather strong guidelines as to how the consistency check of Haberkorn's approach can in principle unfold. 
\subsection{Quantum trajectory approach}
From Haberkorn's theory point of view, it is clear that one cannot agree with possibilities K1 and K2, since these lead to our approach. Hence one has to suggest what is the specific state evolution of surviving RPs, i.e. what is, if any, the state change of radical-pairs until the instant of their recombination into a neutral product.

We will now show that there is limited freedom in doing so. This is because in order to secure consistency in the dynamically simple case $\ks=\kt$, one has to accept what has been until recently the intuitive answer to the previous question, namely that {\it nothing} (besides Hamiltonian evolution) happens to surviving RPs.  As mentioned in Section IIA, the rates $\ks$ and $\kt$ are parameters entering into the master equation, which obviously must be valid for any choice of those parameters. The case $\ks=\kt$ is rather simple dynamically, since in this case RP population decays exponentially at a rate $k\equiv\ks=\kt$, {\it without the decay affecting the state of the surviving RPs}. This means the following. Consider for example a 50/50 mixture of singlet and triplet RPs, having no magnetic interactions (${\cal H}=0$). If $\ks=\kt$, the same number of singlet and triplet RPs will recombine in the interval $dt$, hence at time $t+dt$ the mixture will still be 50/50, albeit having a smaller total population. On the other hand, if $\ks\neq\kt$, the spin character of this mixture would change, becoming more (less) singlet if $\ks<\kt$ ($\ks>\kt$).

To summarize, (i) assuming that having a different fundamental theory for different radical pairs (i.e. different combinations of $\ks$ and $\kt$) is not an acceptable option, and (ii) being unable to propose what happens in the general case to non-recombining RPs from Haberkorn's point of view, except in the special case $\ks=\kt$, where consistency forces one to accept that {\it nothing} else happens besides unitary evolution (to be proved in the following), we take this to be the general answer, and hence Haberkorn's quantum trajectories are formed by the three events presented in Table II.
\begin{table}[h!]
    \centering
    \caption{Quantum trajectory time evolution according to Haberkorn' approach.}
    \begin{tabular}{|c|c|c|}
        \hline
        Name & Event & Probability of event\\\hline\hline
        H1 & singlet recombination & $dr_{\rm S}=\ks dt\langle\psi|\qs|\psi\rangle$\\\hline
        H2 & triplet recombination & $dr_{\rm T}=\kt dt\langle\psi|\qt|\psi\rangle$\\\hline
        H3 & hamiltonian evolution & $1-dr_{\rm S}-dr_{\rm T}$\\\hline
    \end{tabular}
\end{table}

There are two comments to be made. First, the recombination probabilities $dr_{\rm S}$ and $dr_{\rm T}$ are the same in both Haberkorn's approach and ours, since both theories agree in how the singlet and triplet reaction yields, $Y_{\rm j}=\int{dr_{\rm j}}$, with $j={\rm S,T}$, are calculated. Second, we can now easily prove our previous statement about H1-H3 ensuring consistency in the special case $\ks=\kt$. Indeed, taking into account the completeness relation $\qs+\qt=\mathbbmtt{1}$, it follows that when $\ks=\kt\equiv k$, Haberkorn's master equation \eqref{hab} becomes $d\rho/dt=-i[{\cal H},\rho]-k\rho$. Defining $\rho=e^{-kt}R$, it follows that $dR/dt=-i[{\cal H},R]$. It is thus evident that apart from an exponential decay of $\tr\{\rho\}$, the only radical-pair state change is due to the Hamiltonian evolution. In terms of the quantum trajectories H1-H3, we can retrieve the master equation $d\rho/dt=-i[{\cal H},\rho]-k\rho$ as follows. Since $\qs+\qt=\mathbbmtt{1}$, it is $dr_{\rm S}+dr_{\rm T}=kdt\tr\{\rho\}$. The single-RP density matrix is $\rho/\tr\{\rho\}$, hence averaging H1-H3 leads to 
\begin{align}
\rho+d\rho&=dr_{\rm S}(\rho-{\rho\over {\tr\{\rho\}}})\nonumber\\
&+dr_{\rm T}(\rho-{\rho\over {\tr\{\rho\}}})\nonumber\\
&+(1-dr_{\rm S}-dr_{\rm T})(\rho-i[{\cal H},\rho]dt),
\end{align}
from which it easily follows that indeed $d\rho/dt=-i[{\cal H},\rho]-k\rho$. To reiterate, the quantum trajectories formed by H1-H3 {\it exactly} reproduce the master equation \eqref{hab} in the special case $\ks=\kt$. In the following we will show that this consistency check fails in the general and more interesting (in terms of realistic applications) case $\ks\neq\kt$.
\section{Testing the consistency of Kominis' and Haberkorn's approaches using CIDNP observables}
We will now demonstrate that while our approach is largely (but still not perfectly) consistent, Haberkorn's approach is highly inconsistent. We will use a simple one-nuclear-spin radical-ion-pair with parameters relevant to CIDNP experiments. We stress that how many nuclear spins we consider, or which particular Hamiltonian we pick to exhibit the aforementioned inconsistency is of no concern, since as well known, it takes many supporting cases to establish a theory, {\it but just one counterexample to invalidate it}. Nevertheless, we choose a Hamiltonian of the same form considered in CINDP works like \cite{jeschke1998}, 
\beq
{\cal H}={{\Delta\omega}\over 2}s_{Az}-{{\Delta\omega}\over 2}s_{Dz}+\omega_{I}I_{z}+As_{Az}I_{z}+Bs_{Az}I_{x},\label{ham}
\eeq
where $\Delta\omega$ is the difference in the Larmor frequencies of donor and acceptor electrons due to $\Delta g$, $\omega_I$ the nuclear Larmor frequency, and $A$ and $B$ isotropic and anisotropic hyperfine coupling constants. As in \cite{daviso}, we take a magnetic field of 5 T along the z-axis.  For the rest of the parameters we use $\Delta g=4\times 10^{-4}$, $A=\Delta\omega$ and $B=A/2$. Finally, we use the asymmetric recombination rates $\ks=(20~{\rm ns})^{-1}$ and $\kt=(1~{\rm ns})^{-1}$ pertinent to photosynthetic reaction centers \cite{matysik_review}. 
\begin{figure*}[!]
\begin{center}
\includegraphics[width=14 cm]{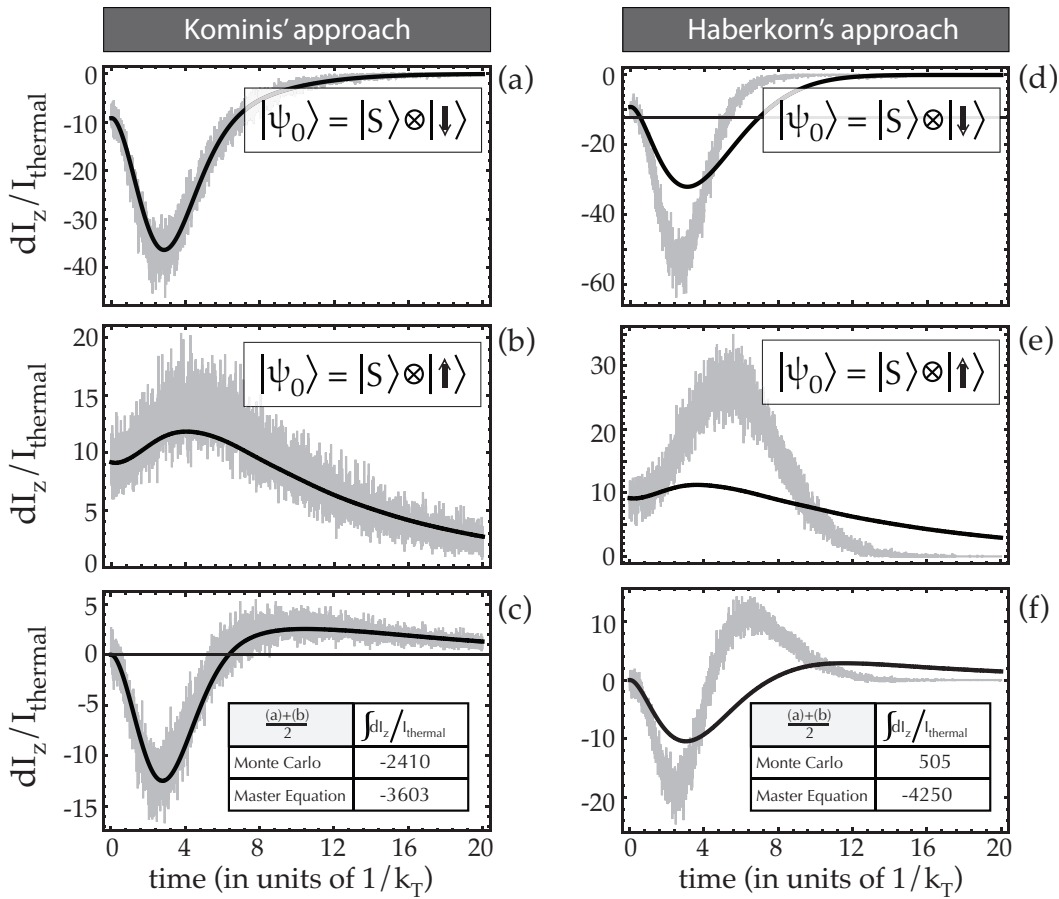}
\caption{Monte Carlo (grey) and master equation (black) calculation of the nuclear spin $dI_z$ deposited to the neutral reaction products as a function of time, normalized by the thermal nuclear spin at the field of 5 T and 233 $^{\circ}$K. (a)-(c) Kominis' and (d)-(f) Haberkorn's approach. We use two different initial conditions, $|\psi_0\rangle=\ket{\rm S}\otimes\ket{\Downarrow}$ (a,d) and $|\psi_0\rangle=\ket{\rm S}\otimes\ket{\Uparrow}$ (b,e). In (c) and (f) we plot the average [(a)+(b)]/2 and [(d)+(e)]/2, respectively, in order to simulate an initially unpolarized nuclear spin. In the insets of (c) and (f) we show the integral of the respective traces, reflecting the total nuclear spin of the reaction products at the end of the reaction.}
\label{dIz}
\end{center}
\end{figure*}

We now calculate the nuclear spin deposited to the reaction products, $dI_z$, as a function of time \cite{JM}.\newline 
{\bf (a) Density matrix propagation} If $\rho$ is the RP density matrix at time $t$, during $dt$ there will be $dr_{\rm S}=\ks dt\tr\{\rho\qs\}$ singlet and $dr_{\rm T}=\kt dt\tr\{\rho\qt\}$ triplet neutral products, the properly normalized density matrix of which 
is $\rho_s=\qs\rho\qs/\tr\{\rho\qs\}$ and $\rho_t=\qt\rho\qt/\tr\{\rho\qt\}$, respectively. Hence the total (singlet + triplet) ground-state nuclear spin accumulated during $dt$ is 
\begin{align}
dI_z&=dr_{\rm S}\tr\{I_z\rho_s\}+dr_{\rm T}\tr\{I_z\rho_t\}\\
&=dt\Big(\ks\tr\{I_z\qs\rho\qs\}+\kt\tr\{I_z\qt\rho\qt\}\Big)
\end{align}
{\bf (b) Quantum trajectories} If $|\psi\rangle$ is the state of the radical-pair at the random instant of recombination, then the properly normalized state of the singlet and triplet reaction product is $|\psi_{s}\rangle=\qs|\psi\rangle/\sqrt{\langle\psi|\qs|\psi\rangle}$ and $|\psi_{t}\rangle=\qt|\psi\rangle/\sqrt{\langle\psi|\qt|\psi\rangle}$, respectively. Hence the nuclear spin deposited to the reaction product is $\langle\psi_{s}|I_{z}|\psi_{s}\rangle$ for a trajectory terminating with a singlet recombination, and $\langle\psi_{t}|I_{z}|\psi_{t}\rangle$ for a trajectory terminating with a triplet recombination.

We evolve the quantum trajectories, and we numerically solve the master equation for two different initial conditions, (I1) $|\psi_0\rangle=\ket{\rm S}\otimes\ket{\Downarrow}$, and (I2) $|\psi_0\rangle=\ket{\rm S}\otimes\ket{\Uparrow}$.  We then average the results of (I1) and (I2) in order to simulate the realistic scenario of starting with an unpolarized nuclear spin. Finally, we integrate the averaged traces in order to find $\int dI_z$, which is directly accessible in CIDNP experiments. This is normalized by the Boltzmann equilibrium nuclear spin, which for the considered parameters is $I_{\rm thermal}\approx 10^{-5}$. 

In Fig.\ref{dIz} we show the main result of this work. The level of inconsistency of Haberkorn's approach is evident just by a visual inspection of Figs.\ref{dIz}(d-f). Even more impressive is the result for $\int dI_z$, reproduced for convenience in Table III. The results of Haberkorn's master equation and Haberkorn's quantum trajectories are not only different in magnitude by 740\%, but are also of different sign. In our approach the two results are of the same sign and different in magnitude by 50\%. 
\begin{table}[t!]
\centering
\caption{Internal consistency of Haberkorn's and Kominis' approaches to radical-pair quantum dynamics tested by the reaction products' integrated nuclear spin polarization relevant to CIDNP measurements.}
 \begin{tabular*}{\hsize}{@{\extracolsep{\fill}}l||cc} \hline${{\int dI_z}/I_{\rm thermal}}$&Kominis&Haberkorn\cr
\hline\hline
Master Equation&-3603&-4250\cr
Monte Carlo&-2410&505\cr
\hline
\end{tabular*}
\end{table}
In Appendix A we describe in detail how the quantum trajectories are produced, while in Appendix B we elaborate on the accuracy of our calculations.
\section{Discussion}
The consistency of a theory is not proof of correctness, but on the contrary, the inconsistency of a theory is proof of its inadequacy. Hence while this work provides a supporting argument that our approach is in the right direction, it unravels a major inconsistency within the traditional approach to radical-pair dynamics. 

Until now, all the interesting information about the physical properties of photosynthetic reaction centers were extracted from CIDNP signals based on the traditional understanding of the radical-pair mechanism. In other words, several mechanisms so far understood to produce enhanced nuclear spin polarizations are mostly based on the combined action of very specific Hamiltonian interactions and radical-pair reactions kinetics. It is clear that the extracted physical information contained in the former will be skewed by the inconsistent description of the latter. 

In the conclusions of \cite{hore} it was stated that "Until an experimental instance is found that requires an alternative description of the recombination kinetics, we recommend continued use of the conventional approach". We believe that a failed consistency check of the conventional theory is quite stronger than an experimental instance challenging the theory. Even more so in light of the previous comment, namely that inconsistent reaction kinetics can contrive with skewed interaction Hamiltonians to produce a deluding agreement with experiments. 
\begin{figure}[t!]
\begin{center}
\includegraphics[width=8.5 cm]{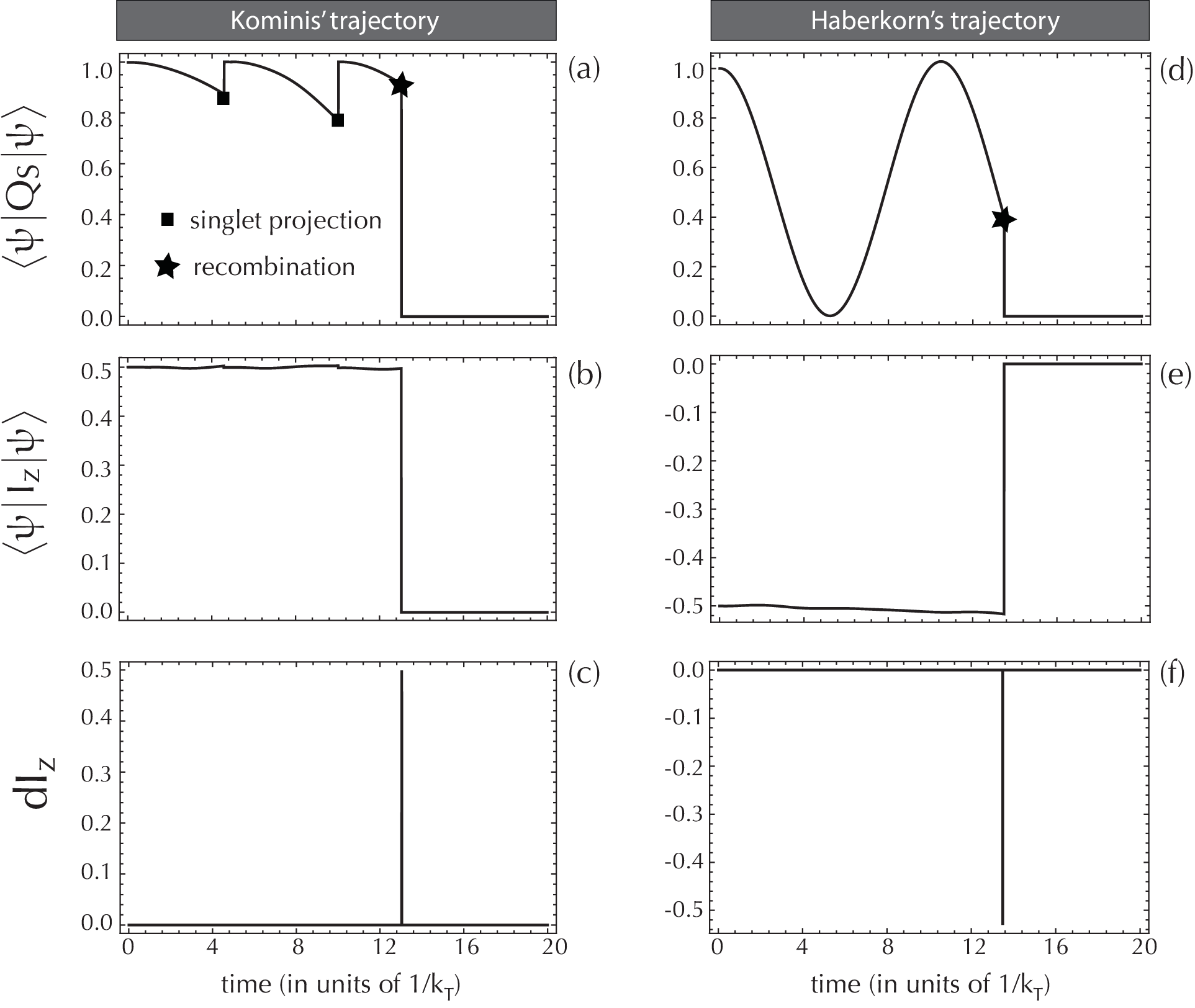}
\caption{(a)-(c) Quantum trajectories in Kominis' approach, the initial state being $\ket{\rm S}\otimes\ket{\Uparrow}$. (d)-(f) Quantum trajectories in Haberkorn's approach, the initial state being $\ket{\rm S}\otimes\ket{\Downarrow}$.  (a,d) singlet probability $\langle\qs\rangle$. Two projections to the singlet state are observed in this particular realization of Kominis' trajectory. Such projections are missing from Haberkorn's trajectories. (b,e) nuclear spin of the radical pair, $\langle\psi|I_z|\psi\rangle$. After the instant of recombination, the trajectory contributes zero to all physical observables. (c,f) nuclear spin $dI_z$ deposited to the reaction products at the instant of recombination. Since the considered radical-pair carries one spin-1/2 nucleus, the magnitude of $dI_z$ correctly is 1/2.}
\label{trajKH}
\end{center}
\end{figure}

To our understanding, singlet and triplet projections are an integral part of the dynamics, the omission of which is responsible for the inconsistent behavior of the conventional theory. The reason is that the Hamiltonian ${\cal V}$ coupling the spin degrees of freedom of the radical-pair to the vibrational reservoir (Fig. 1) is responsible for both effects, random projections and recombination. The former within 2$^{\rm nd}$-order perturbation theory and the latter through 1$^{\rm st}$-order perturbation theory. As shown in \cite{lamb}, both effects are of order ${\cal V}^2$. 

Clearly, there are still unresolved problems that must be addressed. While the quantum trajectory picture of our approach captures what we think are the underlying physics, our master equation does not do a perfect job in matching Monte Carlo. Hence more work is required to address this issue. Lastly, we recommend that the theoretical interpretation of a large number of CIDNP measurements performed over the last several decades should be meticulously revisited in light of a deeper understanding of radical-pair reaction kinetics.
\appendix\section{Generation of quantum trajectories}
For each of the two initial states, we average $2\times 10^5$ quantum trajectories. A single quantum trajectory is generated as follows. We split the time from $t=0$ to $t_{\rm max}=20/\kt$, at which point the reaction is practically over, into $n=5\times 10^3$ steps of duration $dt=t_{\rm max}/n$. We start with the initial state $\ket{\psi_0}$ at time $t=0$, and in each time step we draw a random number $x$ uniformly distributed between 0 and 1. In our approach we calculate the probabilities $dp_{\rm S}$, $dp_{\rm T}$, $dr_{\rm S}$ and $dr_{\rm T}$, and split the real interval [0,1] into five intervals.
If the random number $x$ falls within the
\begin{itemize}\itemsep0.0pt
\item
1$^{\rm st}$ interval of length $dp_{\rm S}$, we realize K1 and move on to the next time step
\item
2$^{\rm nd}$  interval of length $dp_{\rm T}$, we realize K2 and move on to the next time step
\item
3$^{\rm rd}$ interval of length $dr_{\rm S}$, we realize  K3 and terminate the particular trajectory
\item
4$^{\rm th}$ interval of length $dr_{\rm T}$, we realize K4 and terminate the particular trajectory
\item
5$^{\rm th}$ interval of length $1-dp_{\rm S}-dp_{\rm T}-dr_{\rm S}-dr_{\rm T}$, we realize K5 and move on to the next time step.
\end{itemize}
In Haberkorn's approach we calculate the probabilities $dr_{\rm S}$ and $dr_{\rm T}$ and split [0,1] into three intervals. If the random number $x$ falls within the
\begin{itemize}\itemsep0.0pt
\item
1$^{\rm st}$ interval of length $dr_{\rm S}$, we realize  H1 and terminate the particular trajectory
\item
2$^{\rm nd}$ interval of length $dr_{\rm T}$, we realize H2 and terminate the particular trajectory
\item
3$^{\rm rd}$ interval of length $1-dr_{\rm S}-dr_{\rm T}$, we realize H3 and move on to the next time step
\end{itemize} 
Examples of single quantum trajectories are shown in Fig.\ref{trajKH}. For vizualizing the dynamics, we also plot the evolution of the singlet probability, $\langle\qs\rangle$, depicting S-T oscillations. Along those oscillations there are random singlet/triplet projections in Kominis' trajectories, whereas they are completely missing from Haberkorn's trajectories. In both theories, the trajectory is terminated by the recombination event. 
\section{Accuracy of calculations}
The accuracy of the master equation results (first line of Table III) is limited just by the number of time steps $n$. By increasing $n$ beyond $5\times 10^3$, the results converge to numbers within 1\% of the ones stated here.  

In order for the code to run in a practical amount of time, we limited the Monte Carlo simulation (second line of Table III) to $2\times 10^5$ trajectories. The whole simulation (the four Monte Carlo traces of Fig. 2) takes about 9 h in a 4-core machine running at 2.5 GHz. The running time of propagating the density matrix discussed previously is negligible compared to one Monte Carlo trace, since the former involves just one time propagation of an $8\times 8$ matrix, compared to $2\times 10^5$ time propagations of an $8\times 1$ vector required for the latter. 

The accuracy of the integral $\int dI_z$ is easily estimated by taking e.g. 100 consecutive points along a relatively flat part of the traces in Fig.\ref{dIz}, plotting the y-axis values in a histogram, and fitting with a gaussian. The relative error is at the level of 10\%. Thus the positive sign of Haberkorn's Monte Carlo result is correct to within $10\sigma$.

\begin{acknowledgments}
We acknowledge support from the European Union's Seventh Framework Program FP7-REGPOT-2012-2013-1 under grant agreement 316165.
\end{acknowledgments}


\begin{thebibliography}{10}
\bibitem{aspuru}
M. Mosheni, P. Rebentrost, S. Lloyd, A. Aspuru-Guzik, J. Chem. Phys. 129 (2008) 174106.

\bibitem{ishizaki}
A. Ishizaki, G.R. Fleming, Proc. Natl. Acad. Sci. USA 106 (2009) 17255.

\bibitem{scholes}
E. Collini et al., Nature (London) 463 (2010) 644.

\bibitem{mizel}
M.M. Wilde, J.M. McCracken, A. Mizel, Proc. R. Soc. A 466 (2010) 1347.

\bibitem{fleming}
M. Sarovar, A. Ishizaki, G.R. Fleming, K.B. Whaley, Nature Phys. 6 (2010) 462.

\bibitem{scholes_review}
G.D. Scholes, G.R. Fleming, A. Olaya-Castro, R. van Grondelle, Nature Chem. 3 (2011) 763.

\bibitem{brumer}
L.A. Pachón,  P. Brumer, J. Phys. Chem. Lett. 2 (2011) 2728.

\bibitem{engel}
G. Panitchayangkoon et al., Proc. Natl. Acad. Sci. USA 108 (2011) 20909.

\bibitem{plenio}
A.W. Chin, S.F. Huelga, M.B. Plenio, Phil. Trans. Act. Roy. Soc. 370 (2012) 3638 .

\bibitem{rozzi}
C.A. Rozzi et al., Nature Comm. 4 (2013) 1602. 

\bibitem{renger}
T. Renger, F. M\"{u}h, Phys.Chem.Chem.Phys. 15 (2013) 3348.

\bibitem{coker}
P. Huo and D. F. Coker, J. Chem. Phys. 136 (2012) 115102.

\bibitem{collini}
E. Collini, Chem. Soc. Rev. 42 (2013) 4932.

\bibitem{OC}
E. J. O'Reilly, A. Olaya-Castro, Nature Comm. 5 (2014) 3012.

\bibitem{thorwart}
P. Nalbach, C.A. Mujica-Martinez, M. Thorwart, Phys. Rev. E 91 (2015) 022706.

\bibitem{plenio_review}
S.F. Huelga, M.B. Plenio, Contemp. Phys. 54 (2013) 181.

\bibitem{matysik_review}
I.F. C$\acute{e}$spedes-Camacho, J. Matysik in {\it The biophysics of photosynthesis} Goldbeck J, van der Est A (Eds.), Springer Science + Business Media, New York (2014) 141.

\bibitem{closs}
G.L. Closs, L.E. Closs, J. Am. Chem. Soc. 91 (1969) 4549.

\bibitem{kaptein}
R. Kaptein, J.L. Oosterhoff, Chem. Phys. Lett. 4 (1969) 195.

\bibitem{beyerle}
R. Haberkorn, M.E. Michel-Beyerle, Biophys. J. 26 (1979) 489.

\bibitem{boxer}
S.G. Boxer, E.D. Chidsey, M.G. Roelofs, Ann. Rev. Phys. Chem. 34 (1983) 389.

\bibitem{zys}
M.G. Zysmilich, A. McDermott, J. Am. Chem. Soc. 116 (1994) 8362.

\bibitem{mcdermott}
T. Polenova, A.E. McDermott, J. Phys. Chem. B 103 (1999) 535.

\bibitem{prakash}
S. Prakash, P. Gast, H.J.M. de Groot, J. Matysik, G. Jeschke, J. Am. Chem. Soc. 128 (2006) 12794.

\bibitem{diller} 
A. Diller et al., Proc. Natl. Acad. Sci. USA 104 (2007) 12767.

\bibitem{daviso}
E. Daviso et al., J. Phys. Chem. C 113 (2009)10269.

\bibitem{matysik_pnas}
E. Daviso E et al., Proc. Natl. Acad. Sci. USA 106 (2009) 22281. 

\bibitem{matysik_PR}
J. Matysik, A. Diller, E. Roy, A. Alia, Photosynth. Res. 102 (2009) 427.

\bibitem{jeschke_TSM}
G. Jeschke, J. Chem. Phys. 106 (1997) 10072.

\bibitem{JM}
G. Jeschke, J. Matysik, Chem. Phys. 294 (2003) 239.

\bibitem{jeschke_lowB}
G. Jeschke,  B.C. Anger, B.E. Bode, J. Matysik, J. Phys. Chem. 115 (2011) 9919.

\bibitem{haberkorn}
R. Haberkorn, Molec. Phys. 32 (1976)1491.

\bibitem{steiner1}
U. Steiner, T. Ulrich, Chem. Rev. 89 (1989) 51.

\bibitem{steiner2}
K.A. McLauchlan, U.E. Steiner, Molec. Phys. 73 (1991) 241.

\bibitem{ritz2000}
T. Ritz, S. Adem, K. Schulten, Biophys. J. 78 (2000) 707.

\bibitem{woodward}
J.R. Woodward, Prog. React. Kin. Mech. 27 (2002) 165.

\bibitem{rodgers_review}
C.T. Rodgers, Pure Appl. Chem. 81 (2009) 19.

\bibitem{hore_PNAS}
C.T. Rodgers, P.J. Hore, Proc. Natl. Acad. Sci. USA 106 (2009) 353.

\bibitem{pre2009} 
I.K. Kominis, Phys. Rev. E 80 (2009) 056115.

\bibitem{pre2011}
I.K. Kominis, Phys. Rev. E 83 (2011) 056118.

\bibitem{pre2012}
I.K. Kominis, Phys. Rev. E 86 (2012) 026111.

\bibitem{cidnp}
I.K. Kominis, New J. Phys. 15 (2013) 075017.

\bibitem{lamb}
K.M. Vitalis, I.K. Kominis, Eur. Phys. J. Plus 129 (2014) 187.

\bibitem{pre2014}
M. Kritsotakis, I.K. Kominis, Phys. Rev. E 90 (2014) 042719.

\bibitem{petruccione}
H.P. Breuer, F. Petruccione, {\it The theory of open quantum systems}, Oxford University Press, Oxford, UK, 2002.

\bibitem{molin}
V.I. Borovkov, I.S. Ivanishko, V.A. Bagryansky and Y.N. Molin, J. Phys. Chem. A 117 (2013) 1692.

\bibitem{jeschke1998}
G. Jeschke, J. Am. Chem. Soc. 120 (1998) 4425.

\bibitem{hore}
K. Maeda, P. Liddell, D. Gust, P.J. Hore, J. Chem. Phys. 139 (2013) 234309.

\end{thebibliography}
\end{document}